\documentclass{article} % For LaTeX2e
\usepackage[table,xcdraw]{xcolor}
\usepackage{main,times}
\usepackage{textcomp}
\usepackage{amsmath,amssymb,amsfonts}
\usepackage{xspace} %For Spacing
\usepackage{csquotes}
\usepackage{graphicx}
\usepackage{subfig}
\usepackage{enumitem}
\usepackage{multirow}
\usepackage{amssymb}
\usepackage{ifthen}
\usepackage{hyperref}
\usepackage{url}
\usepackage{tikz}
\usetikzlibrary{positioning}
\usepackage{amsmath,amsfonts,bm}

\def\eqref#1{equation~\ref{#1}}
\def\1{\bm{1}}

\DeclareMathAlphabet{\mathsfit}{\encodingdefault}{\sfdefault}{m}{sl}
\SetMathAlphabet{\mathsfit}{bold}{\encodingdefault}{\sfdefault}{bx}{n}

\input{macros}

\title{Perplexed: Understanding When Large Language Models are Confused}

\author{Nathan A. Cooper\thanks{Work done during an internship at ServiceNow} \\
Stability AI \\
\texttt{nathan.cooper@stability.ai} \\
Torsten Scholak  \\
ServiceNow Research \\
\texttt{torsten.scholak@servicenow.com} \\
}
\iclrfinalcopy % Uncomment for camera-ready version, but NOT for submission.
\begin{document}

\maketitle

\begin{abstract}
Large Language Models (LLMs) have become dominant in the Natural Language Processing (NLP) field causing a huge surge in progress in a short amount of time. However, their limitations are still a mystery and have primarily been explored through tailored datasets to analyze a specific human-level skill such as negation, name resolution, \etc In this paper, we introduce \perplexed, a library for exploring where a particular language model is perplexed. To show the flexibility and types of insights that can be gained by \perplexed, we conducted a case study focused on LLMs for code generation using an additional tool we built to help with the analysis of code models called \codetokenizer. Specifically, we explore success and failure cases at the token level of code LLMs under different scenarios pertaining to the type of coding structure the model is predicting, \eg a variable name or operator, and how predicting of internal verses external method invocations impact performance. From this analysis, we found that our studied code LLMs had their worst performance on coding structures where the code was not syntactically correct. Additionally, we found the models to generally perform worse at predicting internal method invocations than external ones. We have open sourced both of these tools to allow the research community to better understand LLMs in general and LLMs for code generation.
\end{abstract}
\section{Introduction}\label{sec:intro}

With the vast quantity of textual data and compute, Large Language Models (LLMs) have rocketed to be the state-of-the-art approaches for modeling natural language (mainly English) text \citep{devlin2018bert, radford2019language, brown2020language}. However, despite their (seemingly) impressive performance, there has been work to show such performance can be quite fragile and have significant limitations \citep{niven2019probing, bender2020climbing}. There has been a surge of research into this area such as \citet{eval-harness}, which introduced a suite of tests designed to evaluate the ability of autoregressive LLMs to handle various tasks such as arithmetic manipulation, high-school and college level tests, and many other tasks requiring various skills. Similarly, \citet{ribeiro2020beyond} explored the ability of LLMs to perform negation, entity replacement, and other simple operations, finding that they struggle with these basic modifications showing a lack of understanding. In addition to these test suite type approaches, the subfield of \emph{Bertology} is focused on probing the types of information present in the intermediate representations of LLMs have also been explore \citep{tenney2019bert, michel2019sixteen, clark2019does, hoover2019exbert, rogers2021primer}. These datasets and probing methods help unravel some of the limitations of theses LLMs and how they work. Another useful lens to view LLMs through is by investigating their failure cases, \ie when an LLM fails for a particular example or set of similar examples.

In this paper, we introduce \perplexed~which allows for the exploration of where a particular LLM is \emph{perplexed} or confused. \perplexed is orthogonal to the above mentioned efforts and can be integrated into such evaluations to give a more holistic view. Specifically, the main difference between our approach is that we focus on the outputs of the model and treat it as a black box instead of probing the internal representations of model making our approach can be more easily applied to different LLM architectures. Additionally, our approach does not require training any probes.

To demonstrate the usefulness of \perplexed, we present a case study focused on LLMs for code generation. Code generation is a challenging task that requires a high level of syntactic and semantic understanding, making it an ideal testbed for evaluating the capabilities and limitations of LLMs \citep{chen2021evaluating, mastropaolo2021studying, xu2022systematic}. In addition to \perplexed, we have also developed a tool called \codetokenizer that helps with the analysis of code models by aligning Byte Pair Encoding (BPE) \citep{sennrich2015neural} tokens with their Abstract Syntax Tree (AST) node\footnote{https://en.wikipedia.org/wiki/Abstract\_syntax\_tree} counterparts. This AST representation is quite useful as every program can be represented by it and it gives a formal and concise representation for understanding different coding constructs, \eg variable identifiers or operators, that are represented in a program. We use \perplexed and \codetokenizer to explore success and failure cases of code LLM models on predicting different coding constructs as well as how external verses external method invocations, \ie calling a third-party library verses calling a method defined somewhere else in the software system, impact performance. From our analysis, we made the following findings:

\begin{enumerate}
    \item The studied code LLMs performed worst on coding structures that were not syntactically correct.
    \item The models generally performed worse at predicting internal method invocations than external ones.
\end{enumerate}

% From this analysis, we found that our studied code LLMs had their worst performance on coding structures where the code was not syntactically correct. Additionally, we found the models to generally perform worse at predicting internal method invocations than external ones. We have open sourced both of these tools to allow the research community to better understand LLMs in general and LLMs for code generation.

Overall, the goal of this paper is to provide researchers with tools that can help them better understand LLMs and their capabilities and limitations. By open sourcing \perplexed\citep{perplexed} and \codetokenizer\citep{codetokenizers} and our findings, we hope to enable the research community to more effectively study LLMs and advance the state of the art in this area.

\section{Perplexed Overview}\label{sec:tool}

\begin{displayquote}
When you sort your dataset descending by loss you are guaranteed to find something unexpected, strange and helpful.

\signed{Andrej Karpathy}
\end{displayquote}

Inspired by Andrej Karpathy's above suggestion to look into the worst performing examples of a model in order to gain a deeper understanding of a model's behavior \citep{andrej2020loss}, we developed \perplexed. \perplexed is a library for more detailed analysis of large language models (LLMs) for text generation. The core idea behind \perplexed is to evaluate a model's performance, in terms of perplexity or cross-entropy, at the per-token level rather than across an entire dataset. This enables a more fine-grained analysis of a model, allowing for greater insight into its strengths and weaknesses.

For example, if you find that your model tends to have high perplexity for the token \emph{esoteric}, you can hypothesize that this is most likely due to the token being rarely seen in the training dataset. You can confirm this hypothesis by counting the number of times the token \emph{esoteric} appears in the training data. One way to improve the model's performance in this case would be to collect more training examples that contain the token \emph{esoteric}.

\begin{figure}[h]
    \centering
    \includegraphics[width=0.2\textwidth]{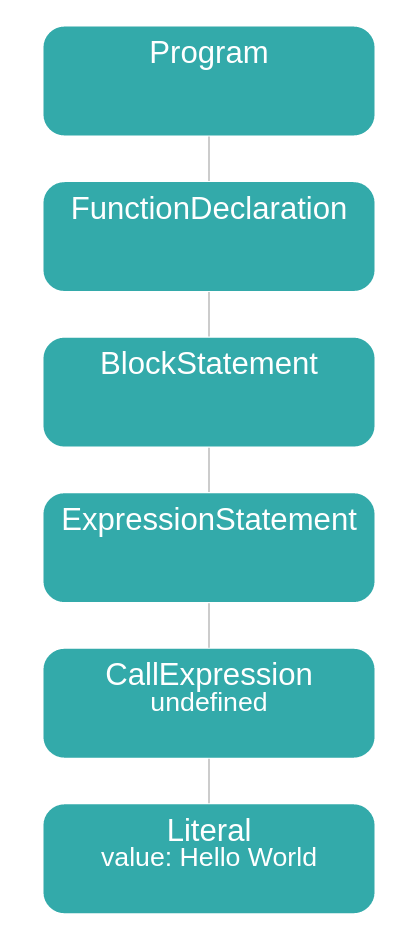}
    \caption{Example of an AST representation for a hello world program}
    \label{fig:ast_example}
\end{figure}

An interesting feature of \perplexed is the ability to align what we call \emph{semantic} tokens with their Byte Pair Encoding (BPE) counter parts. For example, a \emph{semantic} token might represent what part of speech (POS) a particular BPE token is. For example, in the sentence "I ran yesterday," where the BPE token \emph{ran} would be tagged with the verb POS tag. This allows for even more interesting questions to be investigated such as which POS tags does the model struggle with the most. To demonstrate the power of this feature, we created \codetokenizer, a library for aligning Abstract Syntax Tree (AST) nodes to BPE tokens in code data. Every program can be represented in this AST representation and it can be a useful lens to view programs through. Figure \ref{fig:ast_example} gives an example of an AST for a hello world program\footnote{Figure generated using: \url{http://nhiro.org/learn_language/AST-Visualization-on-browser.html}}. With \codetokenizers, a researcher can now easily investigate questions such as which AST node types does a model struggle the most with providing greater insight into the model's limitations.

Using this \emph{semantic} token feature we investigate the limitations of a recent popular LLM for code generation. Specifically, we look at the types of AST nodes it struggles the most with along with the BPE tokens. \textcolor{black}{Additionally, \citet{hellendoorn2019code} showed one of the primary focuses of code completion, a subset of code generation, is on method invocations, specifically internal method invocations where a developer is calling a method defined in the same software system they are developing rather than an external method invocation such as one to a third party library. Therefore, we set out to investigate our studied LLM for code generation on its performance on internal versus external method invocations as well.}

% performance on internal versus external method invocations. We discuss this case study in more depth in the following section.

\subsection{Implementation Details}

\begin{figure}%
    \centering
    \subfloat[\centering Example of using \perplexed to evaluate a model on a dataset.]{{\includegraphics[width=.8\linewidth]{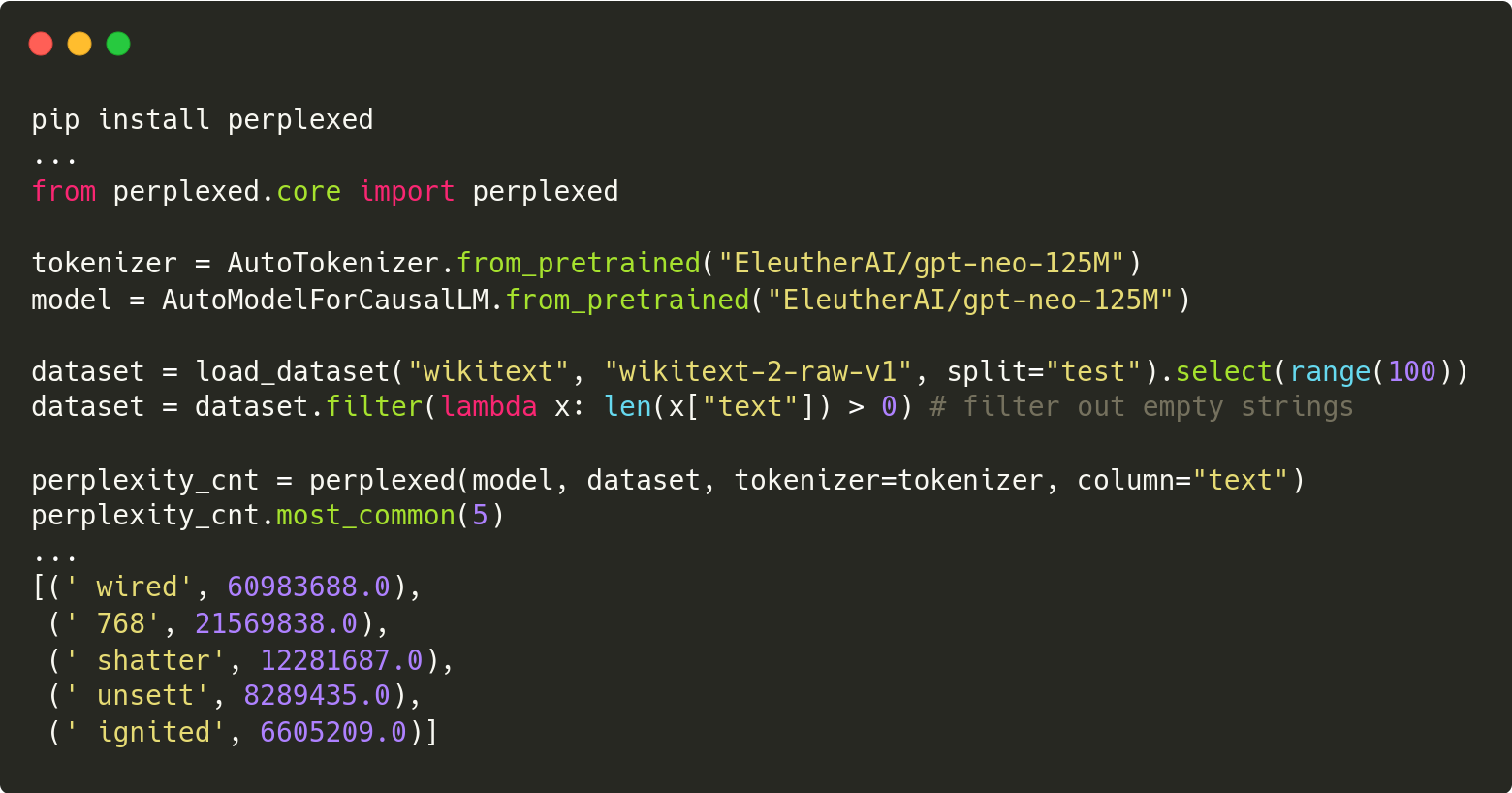} }}%
    \qquad
    \subfloat[\centering Example of using \codetokenizer to tokenize a piece of code and align its AST.]{{\includegraphics[width=.7\linewidth]{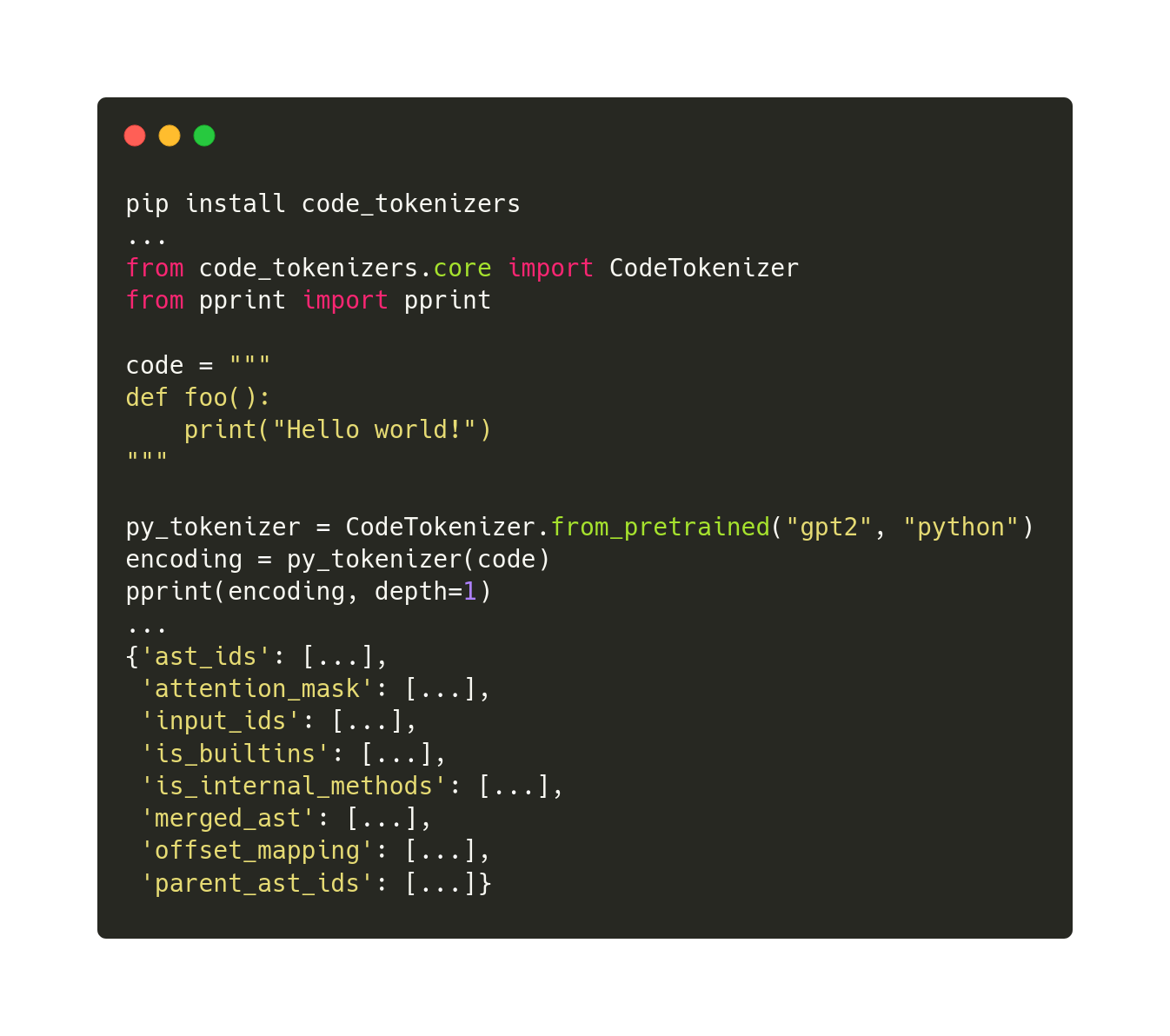} }}%
    \caption{Examples of using \perplexed and \codetokenizer.}%
    \label{fig:examples}
\end{figure}

% \begin{figure}
%   \centering
%   \begin{subfigure}[b]{width=0.5\textwidth}
%     \includegraphics[width=\textwidth]{figs/perplexed.png}
%     \caption{Example of using \perplexed to evaluate a model on a dataset.}
%     \label{fig:perplexity}
%   \end{subfigure}
%   % \\
%   \begin{subfigure}[b]{width=0.5\textwidth}
%     \includegraphics[width=\textwidth]{figs/codetokenizers.png}
%     \caption{Example of using \codetokenizer to tokenize a piece of code and align its AST.}
%     \label{fig:codetokenizer}
%   \end{subfigure}
%   \caption{Examples of using \perplexed and \codetokenizer.}
%   \label{fig:examples}
% \end{figure}

% \begin{figure}[h]
%     \centering
%     \includegraphics[width=\textwidth]{figs/perplexed.png}
%     \caption{Example of using \perplexed to evaluate a model on a dataset.}
%     \label{fig:perplexity}
% \end{figure}

% \begin{figure}[h]
%     \centering
%     \includegraphics[width=0.8\textwidth]{figs/codetokenizers.png}
%     \caption{Example of using \codetokenizer to tokenize a piece of code and align its AST.}
%     \label{fig:codetokenizer}
% \end{figure}

% \begin{figure}%
%     \centering
%     \subfloat[\centering Example of using \perplexed to evaluate a model on a dataset.]{{\includegraphics[width=.8\linewidth]{figs/perplexed.png} }}%
%     \qquad
%     \subfloat[\centering Example of using \codetokenizer to tokenize a piece of code and align its AST.]{{\includegraphics[width=.5\linewidth]{figs/codetokenizers.png} }}%
%     \caption{APIs of \perplexed and \codetokenizer}%
%     \label{fig:api}%
% \end{figure}

Both \perplexed and \codetokenizer are written in Python \citep{python} and integrate into the HuggingFace's \emph{transformers} and \emph{datasets} libraries \citep{transformers, datasets}. Specifically, \perplexed accepts any decoder-only Transformer model along with a dataset to evaluate the model on supported by the \emph{transformers} and \emph{datasets} libraries. Using the \emph{transformers} library allowed for us to retrieve the underlying logit predictions of any decoder-only Transformer model so we can manually calculate the cross-entropy per token as well as have access to a large collection of decoder-only Transformer models through the Huggingface model hub \citep{model_hub}.  Similarly, the \emph{datasets} library also gives us access to a huge variety of datasets to evaluate a model on due to the very popular Huggingface datasets hub \citep{datasets_hub} and a very performant data processing pipelines through the \emph{datasets}' API for things like tokenization of a dataset. Additionally, \perplexed relies on \emph{PyTorch} \citep{pytorch} for performing the cross-entropy calculation. Similarly, \codetokenizer leverages the \emph{PreTrainedTokenizer} class for the BPE tokenization and \emph{tree-sitter} library \citep{treesitter} to generate the Abstract Syntax Tree. To align the two representations, we look at the BPE token and AST node spans in character space to determine when the two are overlapped. Specifically, since multiple BPE tokens can represent a single AST node due to the word the AST node corresponds to such as \emph{if} or \emph{var\_name}, we always look have a one AST node to one or more BPE token relationship ensuring that ending of the first BPE token and the end of the last BPE token is within the span of the word corresponding to the AST node in terms of character indices. For most BPE tokenizers this heuristic works well since they do not allow merging across special tokens such as white space and punctuation that AST nodes do not cross on word spans, \ie white space or the period character dictate a separation between coding structures in the AST parser. However, BPE tokenizers that do not follow this property such as the InCoder \citep{fried2022incoder} tokenizer, which has a single token representing this entire statement "import numpy as np", do not work with \codetokenizer. The interfaces for \perplexed and \codetokenizer can be seen in figures \ref{fig:examples}. Using the literate programming framework \emph{nbdev} \citep{nbdev}, we were able to write clear documentation and tests within the same Jupyter Notebook environment \citep{jupyter}, which greatly accelerated our development process.
\section{Case Study: Analyzing LLMs for Code Generation}\label{sec:case}
In this section, we discuss the study details for showing the types of analysis that \perplexed and \codetokenizers can provide. Specifically, we seek to answer the following research questions:

\begin{enumerate}[label=\textbf{RQ$_{\arabic*}$:}, ref=\textbf{RQ$_{\arabic*}$}, wide, labelindent=5pt]\setlength{\itemsep}{0.2em}
      \item \label{rq:bpe}\textit{Which are the worst and best performing BPE tokens?}
      \item \label{rq:ast}\textit{Which are the worst and best performing AST nodes?}
      \item \label{rq:internal_external}\textit{How is the performance between predicting internal verse external method invocations?}
\end{enumerate}

\textcolor{black}{The first two research questions were chosen since they naturally align with the philosophy behind \perplexed and \codetokenizers, namely, understanding the worst performing predictions and tying them to to code specific concepts. The best predictions are a natural follow up to this philosophy and provide a good contrast between performance generally. The third research question was chosen as it focuses on a common use case of LLMs for code, namely, for the task of code completion, which has been shown to be a struggle for deep learning models \citep{hellendoorn2019code}.}

% http://www.sback.it/publications/icse2019b.pdf

% We chose these particular research questions since 

To answer these research questions, we focus on the recently released \emph{SantaCoder} LLM for code from BigCode \citet{allal2023santacoder}. \emph{SantaCoder} is a $1.1B$ parameter GPT-style model trained on the Java, JavaScript, and Python portions of the Stack \citet{kocetkov2022stack}.

\subsection{Data Collection}
For evaluating \emph{SantaCoder}, we used the \emph{codeparrot/github-code} dataset\footnote{https://huggingface.co/datasets/codeparrot/github-code}, which contains a large collection of code files from GitHub. Specifically, we only used the Python portion that is licensed under GPL-3.0\footnote{https://www.gnu.org/licenses/gpl-3.0.html}. This is because \emph{SantaCoder} was trained exclusively on liberally licensed code \citet{allal2023santacoder}, therefore using only the GPL-3.0 split allows us to have more confidence that the data was not seen during training. This is especially important for \ref{rq:internal_external} since we want to ensure that a specific internal method is not known to the model when determining if the model can perform well when attempting to complete it.

The portion of Python GPL-3.0 licensed code amounted to \originaldata of files. We then took a random $10\%$ of the data since using the full dataset would be too computationally expensive. This resulted in a sample size of \sampledata. We then filtered by the maximum number of characters (\maxlen) since applying the \codetokenizers parser can be slow for very long files resulting in \maxlendata files. Next, since we are concerned with the performance of internal verse external method invocations, we removed any repositories that did not have multiple files in the dataset (\repodata after filtering). Finally, \textcolor{black}{in order to address \ref{rq:internal_external}}, we determined which files had at least one internal method call by extracting all method declarations across the repositories' files and method invocations within a file (ignoring declarations and invocations within the same file) and looked at the intersection of these. If there are some method invocations that intersect with the set of method declaration, we keep the file. This resulted in a final dataset of \finaldata files.

% Since our analysis focuses on the recent SantaCoder model from BigCode \REF{}, we wanted to make sure our evaluation data did not overlap with the training data of SantaCoder. Therefore, we used only GPL-3 licensed code from the github-code huggingface dataset \REF{} since SantaCoder was only trained on liberally licensed code \REF{}. This resulted in a total of \originaldata of python GPL-3 licensed files. We additionally filtered by the maximum number of characters (\maxlen) since applying the \codetokenizers parser can be slow for very long files resulting in \maxlendata files.

% Original Dataset Size: 1,160,889
% Size after Filtering Max Length (4096): 682,267
% Size after

\subsection{Experiments}
To answer our research questions, we evaluated \emph{SantaCoder} using our dataset generated above using \perplexed to capture the most perplexing tokens and \codetokenizers to align the BPE tokens with their AST node counterparts. Specifically, we truncate or pad each document to fit within the max length of the \emph{SantaCoder} model and calculate the loss per token rather than averaging over all tokens as normally done when calculating perplexity. Additionally, we keep all losses for a given token in order to investigate the loss distributions of different tokens. To compare internal versus external method invocations, we filtered out any AST node types that were not associated with a method invocation (\ie we focused on \emph{call} and \emph{argument\_list}). Next, using \codetokenizers we separated out the AST nodes that belonged to an internal method invocation and external method invocation. Next, we remove any method invocations we deem to be too common and not meaningful similar to stop-word removal in Natural Language Processing (NLP) pipelines. Specifically, we remove any method invocations for method names that are the same as Python's builtin methods (\eg print, range, zip, \etc).  Lastly, instead of using the actual perplexity scores, which would be very sparse due to the exponential scale that perplexity applies to the cross-entropies, we use the raw cross-entropies to better visualize the token distributions.

% In addition to internal vs. external method calls, we also looked at the general performance for the top and bottom performing BPE tokens and AST nodes.

\section{Results and Discussion}\label{sec:results}

We will now present and discuss our results.

\subsection{\ref{rq:bpe} Worst and Best BPE Tokens}

\begin{figure}%
    \centering
    \subfloat[\centering Best Performing BPE Tokens]{{\includegraphics[width=.8\linewidth]{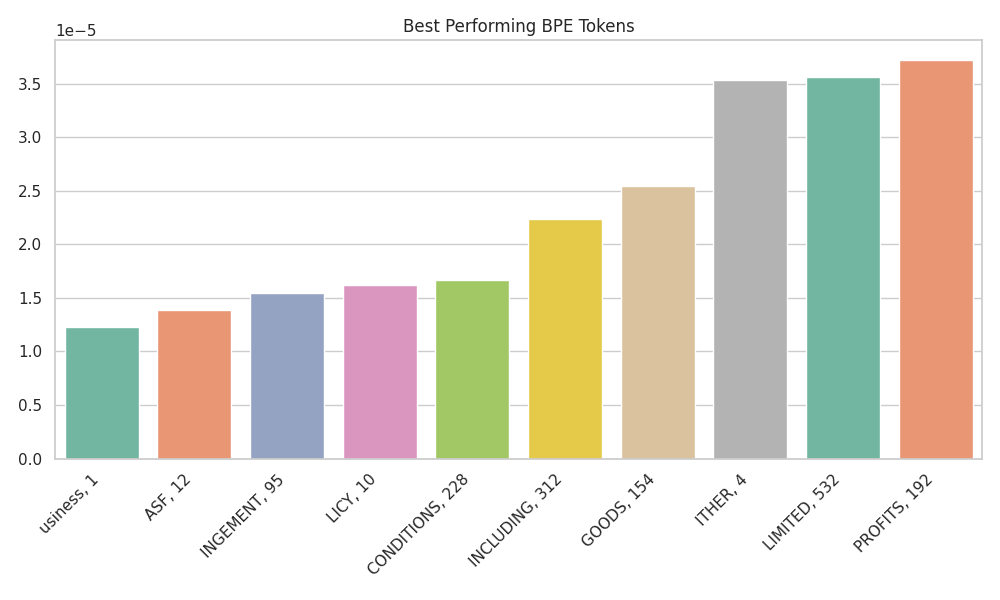} }}%
    \qquad
    \subfloat[\centering Worst Performing BPE Tokens]{{\includegraphics[width=.8\linewidth]{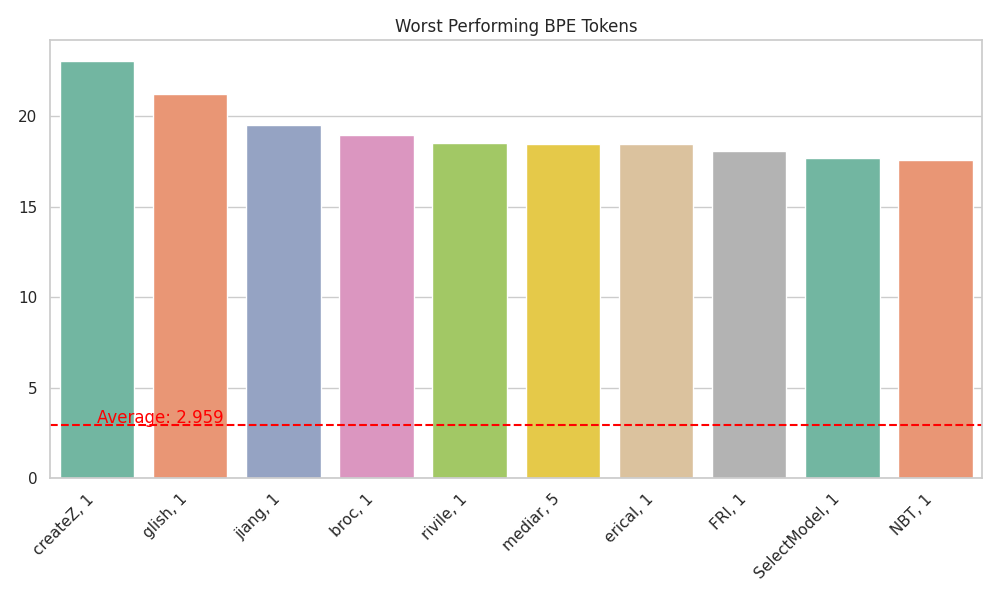} }}%
    \caption{Best and Worst performing BPE tokens with the y-axis being the average cross-entropy and the x-axis being the best or worst performing tokens in terms of their average cross-entropy}%
    \label{fig:bpe_best_worst}%
\end{figure}

Figure \ref{fig:bpe_best_worst} gives an overview of the best and worst performing BPE tokens. Surprisingly many of the top 10 are not heavily represented in our benchmark's dataset with the top token, \emph{usiness}, appearing only once. However, there are many of the top 10 that align with our expectations such as \emph{CONDITIONS} since they most likely belong to the copyright notice of the file. When looking into the worst performing, a trend does appear with nine appearing only once and the other only appearing five times. One hypothesis we wanted to test after viewing these figures was if there is any correlation between token frequency and cross-entropy loss. So, since we have both discrete and continuous values and we are not assuming any type of normality, we used Spearman correlation via Scipy's API\footnote{https://docs.scipy.org/doc/scipy/reference/generated/scipy.stats.spearmanr.html}. From this analysis, we found a relatively strong negative correlation $\approx -0.319$ suggesting that as the token frequency increases the cross-entropy of the token decreases by a small fraction. Intuitively this would make sense as the more common a token is in the dataset, the more times the model has the optimization step to better predict it, but without checking for a causal link it is possible this relationship is spurious.

% Spearman correlation = $-0.3187108402727309$

\subsection{\ref{rq:ast} Worst and Best AST Nodes}

\begin{figure}%
    \centering
    \subfloat[\centering Best Performing AST Nodes]{{\includegraphics[width=\linewidth]{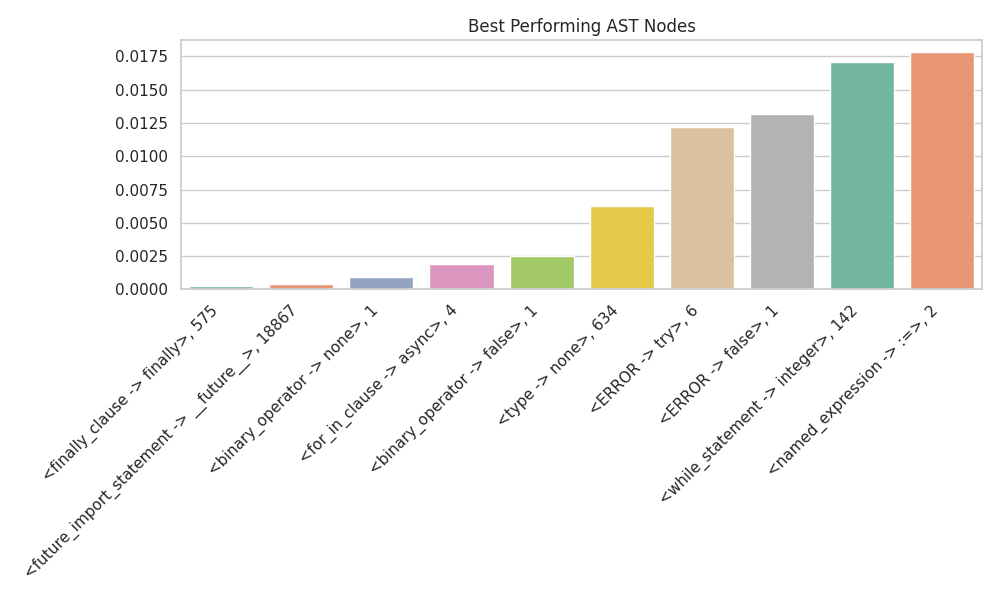} }}%
    \qquad
    \subfloat[\centering Worst Performing AST Nodes]{{\includegraphics[width=.8\linewidth]{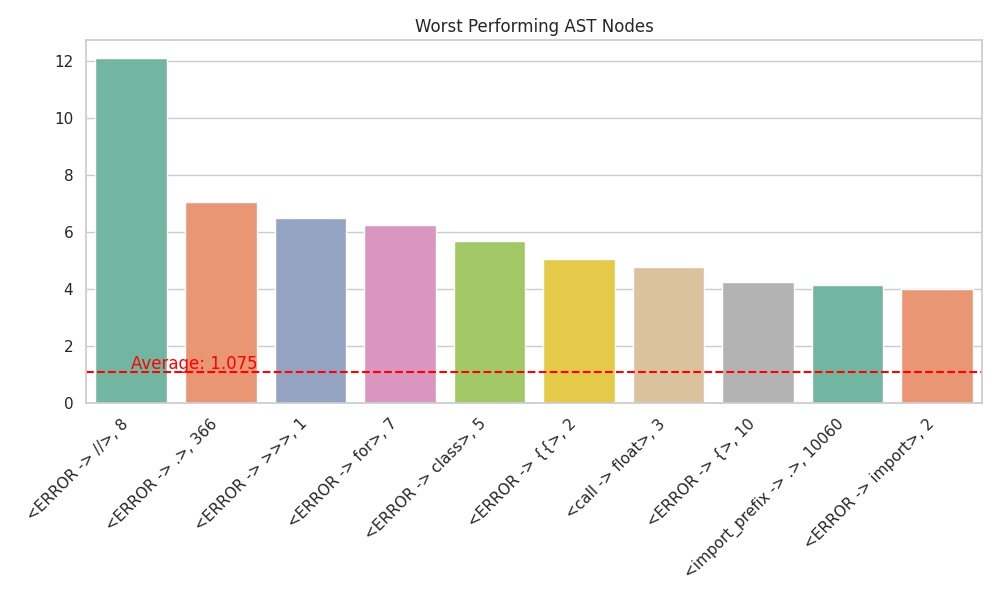} }}%
    \caption{Best and Worst performing AST nodes}%
    \label{fig:ast_best_worst}%
\end{figure}

Similarly, figure \ref{fig:ast_best_worst} shows the best and worst performing AST nodes. In terms of the frequency of the top nodes, there again are a mixture of very infrequent and frequent nodes. Interestingly, there are a few categories that show up in the top 10. Namely, binary and loop operators (both \emph{for} and \emph{while\_statement}), but the binary operators are not the actual operator themselves such as \emph{+/-}, but rather the constants being used in the binary operation (\ie \emph{none} and \emph{false}). This suggests such constructions are relatively easy for the model to predict correctly. When looking at the worst performing nodes a very interesting pattern emerges. Nearly all of the top 10 worst contain the \emph{ERROR} parent node, which represent when the code was not able to be properly parsed due to some syntax error. We take this as strong evidence that our studied LLM for code have a hard time working with improperly written code. Of course, there can be other explanations for such performance such as code with parsing errors is rare so the model is not trained a lot of examples. Therefore, more research is needed to fully conclude this hypothesis. We also performed a similar correlation analysis and found very small correlation of $\approx -0.034$.

% Spearman correlation = $-0.03392354524312661$

\subsection{\ref{rq:internal_external} Internal Vs. External Method Invocation}

% \usepackage{multirow}
\begin{table}[]
\centering
\small
\caption{}
%\vspace{-0.5em}
\label{tabs:internal_external}
\begin{tabular}{cccccc}
\hline
\rowcolor[HTML]{9B9B9B} 
\multicolumn{2}{c|}{\cellcolor[HTML]{9B9B9B}\textbf{Token}}                                                & \textbf{Count} & \multicolumn{1}{c|}{\cellcolor[HTML]{9B9B9B}\textbf{Internal?}} & \multicolumn{1}{c|}{\cellcolor[HTML]{9B9B9B}\textbf{Avg. Cross Entropy}} & \textbf{Std. Cross Entropy} \\ \hline
\multicolumn{1}{c|}{}                                  & \multicolumn{1}{c|}{}                             & 6,236          & \multicolumn{1}{c|}{Yes}                                        & \multicolumn{1}{c|}{0.359}                                               & 1.147                       \\
\multicolumn{1}{c|}{\multirow{-2}{*}{call}}            & \multicolumn{1}{c|}{\multirow{-2}{*}{identifier}} & 297,963        & \multicolumn{1}{c|}{No}                                         & \multicolumn{1}{c|}{0.322}                                               & 1.142                       \\ \hline
\multicolumn{1}{c|}{}                                  & \multicolumn{1}{c|}{}                             & 2,944          & \multicolumn{1}{c|}{Yes}                                        & \multicolumn{1}{c|}{1.215}                                               & 2.201                       \\
\multicolumn{1}{c|}{}                                  & \multicolumn{1}{c|}{\multirow{-2}{*}{(}}          & 675,737        & \multicolumn{1}{c|}{No}                                         & \multicolumn{1}{c|}{0.912}                                               & 1.869                       \\ \cline{2-6} 
\multicolumn{1}{c|}{}                                  & \multicolumn{1}{c|}{}                             & 1,306          & \multicolumn{1}{c|}{Yes}                                        & \multicolumn{1}{c|}{0.690}                                               & 1.343                       \\
\multicolumn{1}{c|}{}                                  & \multicolumn{1}{c|}{\multirow{-2}{*}{)}}          & 335,217        & \multicolumn{1}{c|}{No}                                         & \multicolumn{1}{c|}{0.676}                                               & 1.257                       \\ \cline{2-6} 
\multicolumn{1}{c|}{}                                  & \multicolumn{1}{c|}{}                             & 1,345          & \multicolumn{1}{c|}{Yes}                                        & \multicolumn{1}{c|}{1.187}                                               & 2.131                       \\
\multicolumn{1}{c|}{}                                  & \multicolumn{1}{c|}{\multirow{-2}{*}{,}}          & 197,373        & \multicolumn{1}{c|}{No}                                         & \multicolumn{1}{c|}{0.865}                                               & 1.789                       \\ \cline{2-6} 
\multicolumn{1}{c|}{}                                  & \multicolumn{1}{c|}{}                             & 87             & \multicolumn{1}{c|}{Yes}                                        & \multicolumn{1}{c|}{1.897}                                               & 2.845                       \\
\multicolumn{1}{c|}{}                                  & \multicolumn{1}{c|}{\multirow{-2}{*}{comment}}    & 17,211         & \multicolumn{1}{c|}{No}                                         & \multicolumn{1}{c|}{1.494}                                               & 2.545                       \\ \cline{2-6} 
\multicolumn{1}{c|}{}                                  & \multicolumn{1}{c|}{}                             & 9              & \multicolumn{1}{c|}{Yes}                                        & \multicolumn{1}{c|}{0.234}                                               & 0.623                       \\
\multicolumn{1}{c|}{}                                  & \multicolumn{1}{c|}{\multirow{-2}{*}{false}}      & 2,184          & \multicolumn{1}{c|}{No}                                         & \multicolumn{1}{c|}{0.199}                                               & 0.647                       \\ \cline{2-6} 
\multicolumn{1}{c|}{}                                  & \multicolumn{1}{c|}{}                             & 123            & \multicolumn{1}{c|}{Yes}                                        & \multicolumn{1}{c|}{0.672}                                               & 1.066                       \\
\multicolumn{1}{c|}{}                                  & \multicolumn{1}{c|}{\multirow{-2}{*}{float}}      & 25,448         & \multicolumn{1}{c|}{No}                                         & \multicolumn{1}{c|}{0.879}                                               & 1.271                       \\ \cline{2-6} 
\multicolumn{1}{c|}{}                                  & \multicolumn{1}{c|}{}                             & 5,320          & \multicolumn{1}{c|}{Yes}                                        & \multicolumn{1}{c|}{0.280}                                               & 0.802                       \\
\multicolumn{1}{c|}{}                                  & \multicolumn{1}{c|}{\multirow{-2}{*}{identifier}} & 473,886        & \multicolumn{1}{c|}{No}                                         & \multicolumn{1}{c|}{0.195}                                               & 0.691                       \\ \cline{2-6} 
\multicolumn{1}{c|}{}                                  & \multicolumn{1}{c|}{}                             & 380            & \multicolumn{1}{c|}{Yes}                                        & \multicolumn{1}{c|}{0.738}                                               & 1.149                       \\
\multicolumn{1}{c|}{}                                  & \multicolumn{1}{c|}{\multirow{-2}{*}{integer}}    & 84,628         & \multicolumn{1}{c|}{No}                                         & \multicolumn{1}{c|}{0.512}                                               & 1.031                       \\ \cline{2-6} 
\multicolumn{1}{c|}{}                                  & \multicolumn{1}{c|}{}                             & 23             & \multicolumn{1}{c|}{Yes}                                        & \multicolumn{1}{c|}{0.170}                                               & 0.221                       \\
\multicolumn{1}{c|}{}                                  & \multicolumn{1}{c|}{\multirow{-2}{*}{none}}       & 3,806          & \multicolumn{1}{c|}{No}                                         & \multicolumn{1}{c|}{0.200}                                               & 0.580                       \\ \cline{2-6} 
\multicolumn{1}{c|}{}                                  & \multicolumn{1}{c|}{}                             & 4,851          & \multicolumn{1}{c|}{Yes}                                        & \multicolumn{1}{c|}{1.206}                                               & 2.106                       \\
\multicolumn{1}{c|}{}                                  & \multicolumn{1}{c|}{\multirow{-2}{*}{string}}     & 912,211        & \multicolumn{1}{c|}{No}                                         & \multicolumn{1}{c|}{1.139}                                               & 2.111                       \\ \cline{2-6} 
\multicolumn{1}{c|}{}                                  & \multicolumn{1}{c|}{}                             & 9              & \multicolumn{1}{c|}{Yes}                                        & \multicolumn{1}{c|}{0.675}                                               & 0.709                       \\
\multicolumn{1}{c|}{\multirow{-22}{*}{argument\_list}} & \multicolumn{1}{c|}{\multirow{-2}{*}{true}}       & 2,621          & \multicolumn{1}{c|}{No}                                         & \multicolumn{1}{c|}{0.235}                                               & 0.789                       \\ \hline
\rowcolor[HTML]{9B9B9B} 
\multicolumn{6}{c}{\cellcolor[HTML]{9B9B9B}\textbf{Summary}}                                                                                                                                                                                                                                           \\ \hline
\multicolumn{1}{c|}{Internal}                          & \multicolumn{1}{c|}{}                             & 22,633         & \multicolumn{1}{c|}{}                                           & \multicolumn{1}{c|}{0.777}                                               & 0.493                       \\
\multicolumn{1}{c|}{External}                          & \multicolumn{1}{c|}{}                             & 3,028,246      & \multicolumn{1}{c|}{}                                           & \multicolumn{1}{c|}{0.636}                                               & 0.411                       \\ \hline
\end{tabular}
\end{table}

\begin{figure}%
    \centering
    \subfloat[\centering Internal Method Name Word Cloud]{{\includegraphics[width=.4\linewidth]{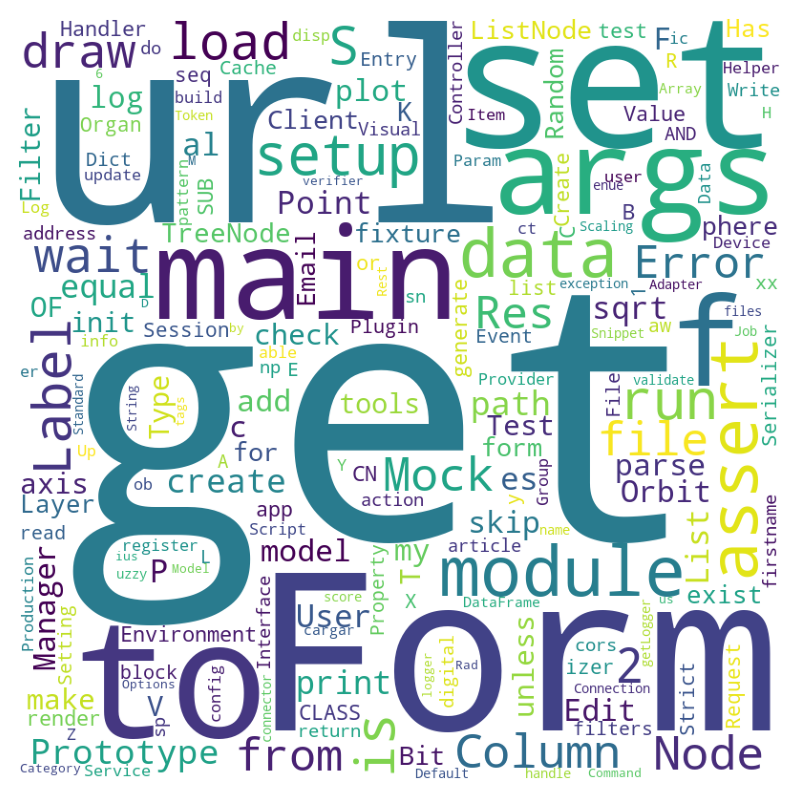} }}%
    \qquad
    \subfloat[\centering External Method Name Word Cloud]{{\includegraphics[width=.4\linewidth]{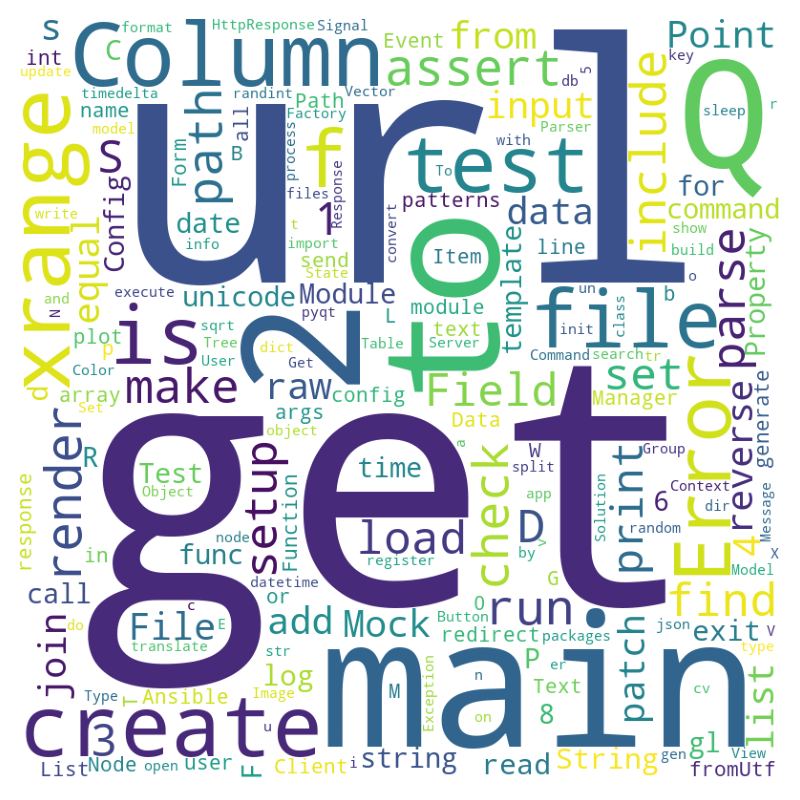} }}%
    \caption{Word Clouds for Internal and External Method Names}%
    \label{fig:word_cloud}%
\end{figure}

Table \ref{tabs:internal_external} gives an overview with the average and standard deviation cross-entropy performance of the various nodes associated with internal and external method invocations (\ie method name and argument list). From this table, you can clearly see a difference ($\approx 0.14B$) between all internal and external method invocation nodes with external being slight easier to predict for \emph{SantaCoder}. This is inline with our hypothesis that external method invocations tend to be more represented in the training data and correspond to things such as popular library usage such as \emph{pandas} or \emph{numpy} \citep{pandas, numpy}. However, we wanted to better understand the exact kind of method names that were represented in both distributions.

Therefore, we additionally created a word cloud of the method names for internal and external method invocations, which you can see in figure \ref{fig:word_cloud}. Please note that the way the \emph{SantaCoder's} BPE tokenizer works, it does not allow for merging across boundary characters such as \emph{`\_'} or \emph{`.'}, so some of the words in the word cloud may be only parts of method invocations due to snake case being the de facto naming convention in the Python ecosystem. As shown, the most prominent BPE tokens in the method name are surprisingly the same (\eg \emph{url} and \emph{get}). However, on closer inspection you are able to see recognizable external methods such as \emph{Path}, \emph{randint}, \emph{pyqt}, \etc Although many are from the Python standard library, we still consider them external since they were not explicitly defined in the software system. Additionally, while many of the names appear to be what should be classified as \emph{builtins}, we believe this to be due to only portions of the name being split by the tokenizer as we do filter out \emph{builtins} before generating the word clouds and any other computations we perform.

Of course, there is always a possibility for misclassification, especially if the code is not syntactically correct. Besides \emph{builtins}, another error in our analysis could be due to not being able to catch all of the methods defined internally as we use automated heuristics to perform this classification based off of the parsed AST. However, we believe our findings align well with hypothesis and we have open sourced all of our data and code for easy reproduction and verification. Additionally, we have a significant test suite for evaluating \perplexed and \codetokenizer. Therefore, we believe our findings are strong evidence towards shedding light on a pitfall of current LLMs for code, namely, the lack of their ability to take in additional context outside of their context window when doing predictions, specifically for method invocations. We hope our analysis and tools can be used to further the research in this area for tackling this problem as well as shed light on other potential issues facing LLMs in general and for code.

\section{Related Work}\label{sec:relwork}

In this section, we will discuss the literature most related to our tools \perplexed and \codetokenizer as well as our case study. Specifically, (i) works and tools for evaluating LLMs, and (ii) works focused on evaluating LLMs for code.

\subsection{LLM Evaluation Works \& Tools}

Evaluating LLMs using tailor made datasets that assess the model's ability to solve a specific task has been a popular method in NLP research \citep{belinkov2019analysis}. SQuAD \citep{rajpurkar2016squad} is one of these types of datasets for evaluating a LLMs ability to tackle the task of reading comprehension (\ie being able to answer a given question based on some context). To evaluate the performance of an LLM, the authors use F1 score of the answers generated by the model. Similarly, WikiQA and TriviaQA are similar reading comprehension Question and Answer (QA) datasets focused on testing an LLMs ability to answer general knowledge questions. Besides QA, other tasks such as summarization have also been investigated \citep{hermann2015teaching, narayan2018don}. While many of these datasets have had LLMs achieve high performance, many works have come out to challenge such progress with datasets designed with adversarial examples that LLMs fail to perform while humans have no issues solving \citep{zellers2019hellaswag, williams2020anlizing, sakaguchi2021winogrande}.

In addition to evaluating LLMs on datasets, tools have been designed to help facilitate model understanding such as CheckList \citep{ribeiro2020beyond}. In CheckList, a user can create templates for evaluating a specific linguistic \emph{capability} such as negation or entity replacement which could then be used to generate a large amount of examples following the created template to evaluate a LLM on. Additionally, \citet{hoover2019exbert} created a tool called \emph{exBERT} for evaluating the hidden representation and attention mechanisms in current LLMs to better understand the inner workings of a model. \perplexed is orthogonal to these efforts in that the focus is on individual predictions of LLMs (\ie next token prediction) and can be combined with a tailor made dataset for a task to give a different lens to see a model's performance through.

% Test sets: , Story Cloze \citet{sharma2018tackling}
% Tools: Eval-Harness \citet{eval-harness}, HF transformers \footnote{https://huggingface.co/docs/transformers/bertology}

\subsection{Evaluating LLMs for Code}

LLMs for code have followed a similar approach to the NLP research field in general with tailor made datasets for evaluating specific task solving abilities such as competitive programming \citep{hendrycks2021measuring, xu2022systematic, li2022competition}, code translation \citep{lachaux2020unsupervised, lu2021codexglue}, or code completion \citep{hellendoorn2017deep, hellendoorn2019code, lu2021codexglue, mastropaolo2021studying}. Competitive programming datasets test a model's ability to solve a wide variety of problem types normally using well known algorithms and data structures and performance is usually measured with functional correctness \citet{hendrycks2021measuring}. Code translation follows in the footsteps of machine translation in NLP measuring the performance of an LLMs ability to translate a piece of code from one programming language to another, which usually is measured with metrics comparing a ground-truth translation using BLEU \citep{papineni2002bleu} or similar metrics designed for code \citep{ren2020codebleu}. Additionally, tools such as those from \citet{lu2021codexglue} and \citet{bigcode-evaluation-harness} combine many of the discussed tasks and metrics into an easy and reproducible test suite for measuring progress in the field.

While \perplexed is not designed specifically with LLMs for code in mind, our case study leveraging our \codetokenizers library shows its potential for such evaluations. However, similar to what was discussed above, \perplexed is orthogonal to related work discussed here. The ability to view a LLM for code's performance on the token level and being able to map that to concepts native to software through the usage of AST nodes opens the door to interesting new research questions.

% Works: Codex \citet{xu2022systematic}, APPs \citet{hendrycks2021measuring}, TransCoder \citet{lachaux2020unsupervised}, CodeXGLUE \citet{lu2021codexglue}, alphacode \citet{li2022competition}

% Tools: BigCode Eval-Harness \citet{bigcode-evaluation-harness}

% \perplexed is a tool to better understand LLM's. The most similar literature to \perplexed is interpretability and specifically a subfield that has evolved to focus on Transformer-based LLM's called \emph{Bertology} \citet{}. We believe \perplexed can help researchers explore interesting questions in the subfield of \emph{Bertology}. Therefore, we discuss the related work of \emph{Bertology} and tools to help facilitate its investigations.

% \textbf{Bertology:}

% Additionally, before Transofmer-based LLM's, there have been extensive research into understanding what type of information is learned by Language Models. 
\section{Conclusion and Future Work}\label{sec:conclusion}

In this paper, we introduced \perplexed, a library for helping understand where a Large Language Model (LLM) is perplexed by analyzing the performance at the per token level. To show the type of analysis \perplexed can provide, we performed a case study of a recent LLM for the task of code generation. To this end, we additionally developed a tool for aligning Byte Pair Encodings (BPEs) \citep{sennrich2015neural} with their Abstract Syntax Tree (AST) node counterparts called \codetokenizer. Through the combination of these tools, we investigated the types of AST nodes LLMs for code struggle with or are most perplexing to the models. Moreover, we also study the difference in performance of an LLM for code on internal verses external method invocations \textcolor{black}{as this has been a common use case of code completion type methods \citep{hellendoorn2019code}}. From our results, we found that our studied LLM for code had nearly all of its top worst performing AST nodes being nodes that had errors due to the code being not syntactically correct. Additionally, we found a difference across all AST nodes associated with method invocations (\ie method name and argument list) showing a worse performance when predicting internal method invocations verse externals, with the only except being the \emph{false} literal used as an argument in the python programming language. Through this evaluation, we shed light on one of the current pitfalls with our studied LLM for code, namely, the lack of additional context from a software system to help with method invocation prediction. We hope by open sourcing both \perplexed and \codetokenizer along with our code and data for easy verification and reproduction that the research community will be more effective in studying LLMs in general and for code generation.

\subsubsection*{Acknowledgments}
We would like to thank Siva Reddy for his invaluable discussion on our approach and feedback on drafts of this paper.

\bibliography{main}
\bibliographystyle{main}

% \appendix
% \section{Appendix}

\end{document}